\documentclass[vecphys]{svmult}

\usepackage{makeidx}         
\usepackage{graphicx}        
\usepackage{multicol}        
\usepackage[bottom]{footmisc}

\makeindex             

\begin{document}
\def\sun{\hbox{$\odot$}}
\def\earth{\hbox{$\oplus$}}
\def\la{\mathrel{\hbox{\rlap{\hbox{\lower4pt\hbox{$\sim$}}}\hbox{$<$}}}}
\def\ga{\mathrel{\hbox{\rlap{\hbox{\lower4pt\hbox{$\sim$}}}\hbox{$>$}}}}
\def\sq{\hbox{\rlap{$\sqcap$}$\sqcup$}}
\def\arcmin{\hbox{$^\prime$}}
\def\arcsec{\hbox{$^{\prime\prime}$}}
\def\fd{\hbox{$.\!\!^{d}$}}
\def\fh{\hbox{$.\!\!^{h}$}}
\def\fm{\hbox{$.\!\!^{\rm m}$}}
\def\fs{\hbox{$.\!\!^{s}$}}
\def\fdg{\hbox{$.\!\!^\circ$}}
\def\farcm{\hbox{$.\mkern-4mu^\prime$}}
\def\farcs{\hbox{$.\!\!^{\prime\prime}$}}
\def\fp{\hbox{$.\!\!^{\scriptscriptstyle\rm p}$}}
\def\micron{\hbox{$\mu$m}}

\newcommand{\etal}{{et~al.}\,}      
\newcommand{\eg}{{e.g.}\,}         
\newcommand{\ie}{{i.e.}\,}         
\newcommand{\cf}{{\it cf.}\,}          
\newcommand{\CDOT}{$\cdot$}             
%
%
\newcommand{\Xray}{\hbox{X-ray}}
\newcommand{\Xrays}{\hbox{X-rays}}
%
%
\def\deg{{^\circ}}
\newcommand{\Deg}{$^\circ$}
\newcommand{\MAG}{$^m\llap{.\thinspace}$}
\newcommand{\Mag}{^m\llap{.\thinspace}}
\newcommand{\rsun}{\,{\rm R}_\odot}
\newcommand{\RSUN}{${\rm R}_\odot$}
\newcommand{\msun}{\,{\rm M}_\odot}
\newcommand{\MSUN}{${\rm M}_\odot$}
\newcommand{\lsun}{\,{\rm L}_\odot}
\newcommand{\LSUN}{${\rm L}_\odot$}
\newcommand{\vsun}{\,v_\odot}
\newcommand{\VSUN}{$v_\odot$}
\newcommand{\kms}{{\,km\,s$^{-1}$}}
\newcommand{\KMS}{{$km\,s^{-1}$}}
\newcommand{\HA}{\mbox{H$\alpha$}}
\newcommand{\LA}{\mbox{Lyman-$\alpha$}}
\newcommand{\NII}{\mbox{\normalsize [N\thinspace\footnotesize II\normalsize ]}}
\newcommand{\NIII}{\mbox{\normalsize [N\thinspace\footnotesize III\normalsize ]}}
\newcommand{\OI}{\mbox{\normalsize [O\thinspace\footnotesize I\normalsize ]}}
\newcommand{\OII}{\mbox{\normalsize [O\thinspace\footnotesize II\normalsize ]}}
\newcommand{\OIII}{\mbox{\normalsize [O\thinspace\footnotesize III\normalsize ]}}
\newcommand{\SII}{\mbox{\normalsize [S\thinspace\footnotesize II\normalsize ]}}
\newcommand{\HI}{\mbox{\normalsize H\thinspace\footnotesize I}}
\newcommand{\HII}{\mbox{\normalsize H\thinspace\footnotesize II}}
\newcommand{\MHI}{$M_{HI}$}
\newcommand{\MT}{$M_T$}
\newcommand{\BT}{$B_T^{0,i}$}
\newcommand{\MBT}{$M_{B_T}^{0,i}$}
\newcommand{\SHI}{$\sigma_{HI}$}
\newcommand{\MHILB}{$M_{HI}/L_B$}
\newcommand{\MHIMT}{$M_{HI}/M_T$}
\newcommand{\MTLB}{$M_T/L_B$}
\newcommand{\DVT}{$\Delta v_{20}$}
\newcommand{\DVF}{$\Delta v_{50}$}
\newcommand{\DVI}{$\Delta v_{0,i}$}
\newcommand{\LDV}{$\log\Delta v_{0,i}$}
\newcommand{\hi}{H\thinspace{\protect\scriptsize I}}
\newcommand{\bi}{\mbox{$B-I$}}
\newcommand{\br}{\mbox{$B-R$}}
\newcommand{\uv}{\mbox{$U-V$}}
\newcommand{\ij}{\mbox{$I-J$}}
\newcommand{\jk}{\mbox{$J-K$}}
\newcommand{\B}{{$B$}}
\newcommand{\V}{{$V$}}
\newcommand{\R}{{$R$}}
\newcommand{\II}{{$I_c$}}
\newcommand{\J}{{$J$}}
\newcommand{\HH}{{$H$}}
\newcommand{\K}{{$K_s$}}
\newcommand{\tfr}{Tully\,--\,Fisher relation}
\newcommand{\ccd}{colour\,--\,colour diagram}
\newcommand{\ML}{\mbox{${\cal M/L}$}}

%
\def\apj    {{ApJ }}
\def\apjl   {{ApJL }}
\def\apjs   {{ApJS }}
\def\aj     {{AJ }}
\def\aaa    {{A\&A }}
\def\aal    {{A\&AL }}
\def\aas    {{A\&ASS }}
\def\apss   {{ApSS }}
\def\apsss  {{ApSSS }}
\def\aplc   {{ApLC }}
\def\araa   {{ARA\&A }}
\def\aarv   {{A\&ARv }}
\def\baas   {{BAAS }}
\def\mnras  {{MNRAS }}
\def\nat    {{Nature }}
\def\rmaa   {{RvMxAA }}
\def\rmaacs {{RvMxAACS }}
\def\pasa   {{PASA }}
\def\pasp   {{PASP }}
\def\pasj   {{PASJ }}
\def\sci    {{Science }}
\def\sal    {{Sov. Astron.\ Lett. }}
\def\paz    {{Pis'ma Astron.~Zh. }}
\def\aspcs  {{ASP Conf. Ser. }}

\title*{Outlining the Local Void with the Parkes HI ZOA and Galactic 
Bulge Surveys}
\titlerunning{Outlining the Local Void in HI} 
\author{Ren\'ee C. Kraan-Korteweg\inst{1},
Nebiha Shafi\inst{1},
Baerbel Koribalski\inst{2},
Lister Staveley-Smith\inst{3},
Peter Buckland\inst{3},
Patricia A. Henning\inst{4}\and
Anthony P. Fairall\inst{1}}
\authorrunning{Kraan-Korteweg et al.} 
\institute{Astronomy Department, University of Cape Town, SA
\texttt{kraan@circinus.ast.uct.ac.za}
\and Australian Telescope National Facility, CSIRO, Epping, Au
\and School of Physics, University of Western Australia, Au
\and Institute for Astrophysics, University of New Mexico, Albuquerque, USA}


%
%
\maketitle

\section{The Local Void}
\label{LV}

The Local Void (LV) was first identified by Tully \cite{tul87} as a
very local region ($cz \la 3000$\kms) devoid of galaxies located next
to the Galactic Bulge $0\deg \la \ell \la 90\deg$ within Galactic
latitudes of $|b| \la 30\deg$. The lack of galaxies in that area must
partly be influenced by the high foreground extinction and star
density. Nevertheless, the reality of the void was never doubted as it
extends beyond the optical and near-infrared Zone of Avoidance (ZOA)
and can be traced to relatively high latitudes (see Figs.~\ref{wedge}
and \ref{slices}). The continuity of low galaxy density across the
opaque part of the ZOA was recently corroborated by HIPASS
\cite{kor04,mey04,won06}, which is unaffected by extinction or star
density.

The actual size, extent and so-called ``emptiness'' of the Local Void
has remained a matter of much debate. Some suggest the LV to be larger
and extend to the more distant Microscopium (or Sagittarius) Void at
$cz \sim 4500$\kms\ (e.g. \cite{fai98}), while others identified
filamentary structures at larger longitudes ($\ell \sim 60\deg$) that
separate off two distinct smaller voids (Delphinus and Cygnus
\cite{nak97,fai98,don05}), hence reducing the size of the
LV. Independently, careful scrutiny of sky survey plates in
Hercules/Aquila did reveal various dwarf galaxies \cite{kar99}
suggesting that the LV might not be quite as galaxy-free as previously
thought.

The interest in this nearest of voids recently escalated again with
the claims on dynamical grounds, that the LV must be much larger (of
the order of 50Mpc) and extremely empty -- if not filled with Dark
Energy -- to explain the repulsive peculiar motion of the LV on the
Milky Way of $\sim 260$\kms\ with respect to the Local Supercluster
restframe \cite{tul07,tul07a}. In addition to that, the large peculiar motion
recently found for the dwarf galaxy ESO461-36 in the LV seems to
suggest that it is being catapulted out of the Local Void with
$\sim~230$\kms\ \cite{tul07}.

\section{The Parkes ZOA HI Surveys}
\label{ZOA_HI}

Better and tighter observational constraints on the size and census of
the Local Void can only be achieved through deep HI surveys.  Three
systematic deep HI surveys performed with the Multibeam Receiver of
the 64\,m Parkes radio telescope cover a large fraction of the
LV. Their integration time is a factor $4 - 5$ longer than HIPASS
\cite{mey04,won06} and reach sensitivities of $rms \sim 6$\,mJy
\cite{don05,kra05,kra05a}. The instantaneous velocity range is $-1200$
to $12\,700$\kms\ as in HIPASS. These are the deep southern ZOA survey
(ZOA) \cite{kra05,hen05} plus northern extension (NE) \cite{don05} which
encompass the Galactic longitude range $196\deg \le \ell \le 52\deg$
for latitudes of $|b| \le \pm 5\deg$, plus a recent extension towards
higher latitudes in the Galactic Bulge region, made because of the
interest in large-scale structures such as the Local Void, the Great
Attractor and the Ophiuchus (super)cluster.

The Galactic Bulge extension (GB) has an average of 20 scans, hence
slightly lower compared to the 25 scans of the ZOA and NE surveys.
It extends to Galactic latitudes of $\pm10\deg$ for the longitude
range $332\deg \le \ell \le 36\deg$, reaching up to higher positive
latitudes ($+15\deg$) for $348\deg \le \ell \le 20\deg$. The 
combined survey area is outlined in Fig.~\ref{slices}.

\section{Outlining the Local Void}

\begin{figure}
\begin{center}
\includegraphics[scale=0.65]{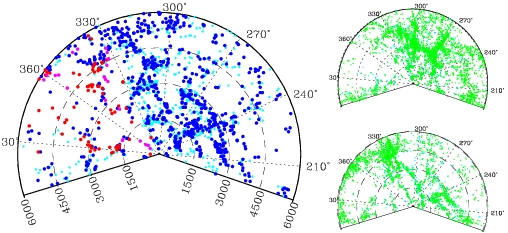}
\caption{Distribution of HI galaxies along the ZOA out to 6000\,km/s
detected in the deep Parkes HI ZOA and NE surveys ($|b| \le 5\deg$;
blue), the GB extension ($\pm10\deg$ resp. $+15\deg$ for $332\deg \le \ell
\le 36\deg$, and $348\deg \le \ell \le 20\deg$) as well as galaxies
from HIPASS (shallower) for $-10\deg \le b \le 15\deg$ (small lighter
dots). The boundaries of the LV lie somewhere within $~ 330\deg \la \ell
\la 60\deg$. For visualisation of continuation of LSS, pie diagrams
above (top) and below (bottom) the GP ($5\deg < |b| < 30\deg$) based
on redshift data in the literature (LEDA) are shown on the right.}
\label{wedge}
\end{center}
\end{figure}

The distribution of the galaxies detected in the deep HI surveys are
displayed in Fig.~\ref{wedge} in a redshift cone out to 6000\kms\
(ZOA+NE with dark (blue) large dots; GB red (lighter) large dots for
$5\deg < |b| < 10\deg$ resp. $+15\deg$ 
including galaxies from the shallower HIPASS for $-10\deg < b
< 15\deg$ (smaller dots; cyan). The redshift cones on the right-hand
side reflect the galaxy distribution above (top) and below (bottom)
the ZOA, for $5\deg < |b| < 30\deg$ respectively, based on published
redshifts as in LEDA.

\begin{figure}
\begin{center}
\includegraphics[scale=0.65]{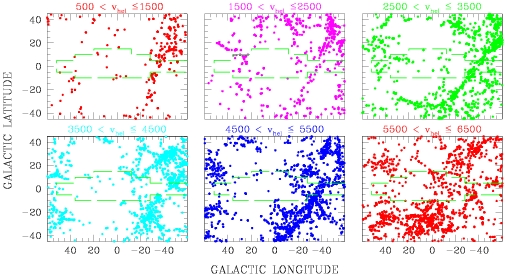}
\caption{Six redshift slices of widths $\Delta v = 1000$\kms\ ranging from 
$500-1500$\kms\ to $5500-6500$\kms\ centered on the GB, respectively the LV
based on the deep Parkes HI surveys (ZOA+NE+GB; the combined HI survey area 
is marked) as well as published redshifts in the literature (LEDA).}
\label{slices}
\end{center}
\end{figure}

When discussing the various features of the LV, the reader is also
referred to Fig.~\ref{slices} which illustrates sky projections (in
gal. coordinates) of redshift intervals of widths $\Delta v =
1000$\kms.
Note that the data outside the outlined HI survey region depends on 
availability of redshift data in  the literature (LEDA) and constitutes
an uncontrolled data set.
  
While the galaxy density in the longitude range of the LV ($\sim
330\deg - 45\deg$) is clearly lower than the rest of the sky
(Fig.~\ref{wedge}), it is by no means as devoid of galaxies as
previously thought. It is in fact quite striking that the LV area
based purely on the deep HI data set (left) -- which is sensitive to
low-mass (gas-rich) dwarf galaxies -- is much more populous compared
to the LV region at higher latitudes (right panels) for which no deep
HI survey is available.

The overall under-dense region seems to extend to about 6000\kms\ (see
both Figs.~\ref{wedge} \& \ref{slices}), hence supportive of Tully's
larger void.  However, a filament is visible around 3000\kms\ which
clearly seems to divide this larger under-dense region into a nearer
and distant void (previous LV and Microscopium Void). But neither of
these two voids appear well defined nor really empty. Within the
nearer LV there seems to exist a protrusion into that void at about
1500\kms (prominent in both Fig.~\ref{wedge} \& \ref{slices}), while an
extension that crosses the Great Attractor Wall at the location of the
Norma cluster seems to cut into the more distant Microscopium void at
$\ell \sim 350\deg, v \sim 4500$\kms\ and $340\deg,5000$\kms. 

It can be argued that the larger LV consists of a huge under-dense
region out to $cz \la 6000$\kms\ from about $345\deg -45\deg$ in
longitude and $-30\deg$ to about $+45\deg$ in latitude, which possibly
is connected to the Cygnus and Delphinus voids at slightly higher
longitudes (see e.g. \cite{don05}). However, various filamentary
features criss-cross and divide this larger LV into smaller voids. And
these smaller voids are not empty either. Even within the nearest part
of the LV ($cz \la 1500$\kms) some -- very low mass -- galaxies have been
detected with the HI surveys.

A project has begun to systematically study the global properties of
the galaxies in and around these voids based on their HI-masses and
near-infrared morphology and luminosity. The latter is being obtained
with the InfraRed Survey Facility (IRSF) at the SAAO (instantaneous
$JHK_s$ bands). Preliminary inspection of some of the LV galaxies
seem to suggest that the galaxies in the voids are extreme low-mass,
faint galaxies while the ones making up the borders of the smaller
voids seem more consistent with normal luminous spiral galaxies.  \\

{\bf Acknowledgements.}  Financial support from the National Research
Foundation as well as the University of Cape Town is kindly
acknowledged (RKK, NS, AF). This research used the Lyon-Meudon
Extragalactic Database (LEDA), supplied by the LEDA team, CRAL,
Obs.~de Lyon.

%
%

\begin{thebibliography}{99}

\bibitem{don05}
Donley J.L., Staveley-Smith L., Kraan-Korteweg, R.C. et al.:
AJ 129, 220 (2005)

\bibitem{fai98}
Fairall A.P.:
\textit {Large-Scale Structures in the Universe},
(Wiley, Chichester 1998)

\bibitem{hen08}
Henning, P.A. et al.: in prep.

\bibitem{kar99}
Karachentseva V.E., Karachentsev I.D., Richter, G.M.:
A\&AS 135, 221 (1999)

\bibitem {kor04}
Koribalski B.S., Staveley-Smith L., Kilborn V.A. et al.:
AJ 128, 16 (2004)

\bibitem{kra05}
Kraan-Korteweg, R.C.:
in \textit{From Cosmological Structures to the Milky Way}, 
RvMA 18, ed. S. R\"oser, (Wiley, New York 2005) pp 48--75

\bibitem{kra00} 
Kraan-Korteweg R.C., Lahav, O.: 
A\&AR, 10, 211 (2000)

\bibitem{kra05a}
Kraan-Korteweg R.C., Staveley-Smith L., Donley J., Koribalski B.,
Henning, P.A.:
in \textit{Maps of the Cosmos}, IAU Symp. 216,
eds. M. Colless, L. Staveley-Smith, \& R. Stathakis, (ASP, San Francisco 2005) 
pp 203--210

\bibitem{mey04} 
Meyer M.J., Zwaan M., Webster R.L. et al.: 
MNRAS 350, 1195 (2004)

\bibitem{nak97}
Nakanishi K., Takata T., Yamada T. et al.:
ApJS 112, 245 (1997)

\bibitem{tul87} Tully R.B., Fisher J.R.:  
\textit{Nearby Galaxies Atlas}, (Cambridge Univ. Press 1987)

\bibitem{tul07} 
Tully, R.B., Shaya, E.J., Karachentsev I.D. et al.:
(astro-ph/0705.4139) 

\bibitem{tul07a} 
Tully, R.B.: these proceedings
(astro-ph/0705.2449) 

\bibitem{won06}
Wong O.I., Ryan-Weber E.V., Garcia-Appadoo D.A. et al.: 
MNRAS 371, 1855 (2006)

\end{thebibliography}
%



\printindex
\end{document}